\documentclass[12pt]{article}
\usepackage[]{}

\date{}
\def\build#1_#2^#3{\mathrel{\mathop{\kern 0pt#1}\limits_{#2}^{#3}}}
\begin{document}

\title{THE TWO-COMPONENT NON-PERTURBATIVE 
POMERON AND THE G-UNIVERSALITY}

\author{Basarab NICOLESCU \\ 
\\
\\
\\
	LPNHE\thanks{Unit\'e de Recherche des Universit\'es Paris 6 et 
		Paris 7, Associ\'ee au CNRS} - LPTPE, Universit\'e Pierre et Marie 
		Curie,\\ 4, Place Jussieu, 75252 Paris Cedex 05, France\\
		nicolesc@in2p3.fr}

\maketitle
\textsl{Talk at the International Workshop on Diffraction in High-Energy 
Physics, Cetraro, Italy, September 2-7, 2000 (to be published in the 
Proceedings of this Conference, Nucl. Phys. B)}
\abstract{
In this communication we present a generalization of the 
Donnachie-Landshoff model inspired by the recent discovery of a 
2-component Pomeron in LLA-QCD by Bartels, Lipatov and Vacca. In 
particular, we explore a new property, not present in the usual Regge 
theory - the \textit{G-Universality} - which signifies the 
independence of one of the Pomeron components on the nature of the 
initial and final hadrons. The best description of the $\bar pp,\ 
pp,\ \pi^\pm p,\ K^\pm p,\ \gamma\gamma$ and $\gamma p$ forward data 
is obtained when G-universality is imposed. Moreover, the $\ell 
n^{2}s$ behaviour of the hadron amplitude, first established by 
Heisenberg, is clearly favoured by the data.}

\vfill
\noindent LPNHE 00-13\hfill October 2000

\newpage


The Donnachie-Landshoff model \cite{Don84} - denoted as $Xs^\epsilon$ in 
the following is very successful in describing $\sigma_{T}$ and 
forward $(t=0)$ $\rho$
data for $\bar pp,\ pp,\ \pi^\pm p,\ K^\pm p,\ \gamma\gamma$ and $\gamma p$
scatterings
 : $\chi^{2}/dof=1.020$ for 16 parameters, 383 data points and 
$\sqrt{s}\geq 9$ GeV \cite{PDG}.

In the present communication I will explore a QCD-inspired 
generalization of this model. The results are obtained in 
collaboration with P. Gauron \cite{GN}.

Recently, Bartels, Lipatov and Vacca \cite{Bar00} discovered the existence 
of a 2-component Pomeron in LLA. The first component is associated 
with 2-gluon exchanges and corresponds to an intercept
\begin{equation}
    \alpha_{P}^{2g}\geq 1.
        \label{1}
\end{equation}
The second component is associated with 3-gluon exchanges with $C=+1$ 
and corresponds to an intercept
\begin{equation}
        \alpha_{P}^{3g}= 1.
        \label{2}
\end{equation}
This last component is exchange-degenerate with the 3-gluon $C=-1$ 
Odderon. 
It is therefore useful to explore possible 2-component Pomeron 
generalizations of the 1-component $X s^{\epsilon}$ Pomeron 

\begin{equation}
        \sigma_{AB}(s)=Z_{AB}+X_{AB}(s)+Y_{AB}^{+}s^{\alpha_{+}-1}
        \pm Y_{AB}^{-}s^{\alpha_{-}-1},
        \label{3}
\end{equation}
where $\sigma_{AB}(s)$ are total cross-sections,
\begin{eqnarray}
        X_{AB}(s) & = &
    X_{AB}s^{\alpha_{P}-1}\label{syst} \label{4} \\
       & = & X_{AB}\ell n\ s \label{5} \\
       & = & X_{AB}\left[
        \ell n^{2}\left(\frac{s}{s_{0}}\right)-
        \frac{\pi^{2}}{4}\right],\label{6}
\end{eqnarray}
and $\alpha_{P},\ \alpha_{+}$ and $\alpha_{-}$ are Reggeon 
intercepts ; $Z_{AB},\ X_{AB},\ Y_{AB}^{\pm},\ s_{0}$ are constants. The + 
sign in front of the $Y_{AB}^-$ term in eq. (\ref{3}) corresponds to 
$\{A=\bar p,\ \pi^-,\ K^-,\ B=p\}$ and the - sign to
$\{A= p,\ \pi^+,\ K^+,\ B=p\}$. If $A=\gamma$ in eqs. (\ref{3})-(\ref{6}), then 
$B=\gamma, \ p$ and $Y^-_{AB}=0$. An implicit scale factor of 1 
(GeV)$^{2}$ is present in the Reggeon and $\ell n\ s$ terms.

The first model in eqs. (\ref{4})-(\ref{6}) - denoted as $Z+Xs^\epsilon$ in the 
following - corresponds to a generalized Donnachie - Landshoff 
approach \cite{GN00,Ko} ; the second - denoted as $Z+X\ell n\ s$ -
to the well known dipole approach \cite{Des} ; the third - denoted as 
$Z+X\ell n^{2}\ s$ - to the Heisenberg - Froissart - Martin form first 
considered in 1952 by W. Heisenberg \cite{Hei65}. The $\rho$-parameter 
is calculated from (\ref{3}) by using the known $s\to se^{-i\pi/2}$ 
crossing rule.

We study, in particular, the following properties :
\begin{itemize}
        \item  [1.]
        The G-universality \cite{GN00,Dak99} ("G" from "gluon") expressed by
        (see eqs. (\ref{3})-(\ref{6}))
\begin{equation}
        X_{AB}(s)=X(s),
        \label{7}
\end{equation}
i.e. independence of $X_{AB}$ on $A$ and $B$ ($A,\ B$= hadrons only), a property 
not present in the usual Regge theory.

        \item  [2.]
        The weak exchange-degeneracy
\begin{equation}
        \alpha_{+}=\alpha_{-},\qquad Y_{AB}^{+}\neq Y_{AB}^-.
        \label{8}
\end{equation}
\end{itemize}

The results for the simultaneous description of 
$\bar pp,\ pp,\ \pi^\pm p,\ K^\pm p,\ \gamma\gamma$ and $\gamma p$ 
reactions are given in Tables I (fits of $\sigma_{T}$ data only) and 
II (fits of $\sigma_{T}$ and $\rho$ data).

\begin{table*}[h]
\caption{Results of the fits of $\sigma_{T}$ data. The symbol = in 
the $\alpha_{+}$ column means weak exchange-degeneracy ($\alpha_{+}=\alpha_{-}$).}
\label{table:1}
\renewcommand{\arraystretch}{1.2} 
{\small 
\begin{tabular}{ccccccccc}
\hline
        & \multicolumn{2}{c}{G-}   
        & \multicolumn{2}{c}{exchange}   & & &  &  \\
  Model &  \multicolumn{2}{c}{Universality}  &  \multicolumn{2}{c}{degeneracy}  
         & $\alpha_{+}$ & $\alpha_{-}$ & $N_{par}$ & $\chi^{2}/dof$  \\ 
        \cline{2-5}
         & Yes & No & Yes & No &  &  &  &   \\
        \hline
        $Xs^{\epsilon}$ &  & x &  & x & 0.66$\pm$ 0.02& 0.45$\pm$ 0.02 & 16 
        & 0.931  \\        
         &  & x & x &  & = & 0.48 & 15 & 1.009  \\
        \hline  
        &  & x &  & x & 0.618$\pm 0.021$ & 0.465$\pm 0.021$ & 17 
        & 0.936  \\
        $Z+Xs^\epsilon$ &  & x & x &  & = & 0.491$\pm 0.023$ & 16 & 0.980  \\
         & x &  &  & x & 0.526$\pm 0.029$ & 0.479$\pm 0.023$ & 17 & 0.835  \\
         & x &  & x &  & = & 0.487$\pm 0.023$ & 16 & 0.836  \\
        \hline
        &  & x &  & x & 0.826$\pm 0.013$ & 0.468$\pm 0.022$ & 16 
        & 0.865  \\
       $Z+X\ell n\ s$  &  & x & x &  & = & 0.586$\pm 0.019$ & 15 & 1.281  \\
         & x &  &  & x & 0.658$\pm 0.007$ & 0.485$\pm 0.022$ & 16 & 1.066  \\
         & x &  & x &  & = & 0.610$\pm 0.016$ & 15 & 1.286  \\
        \hline
        &  & x &  & x & 0.653$\pm 0.026$ & 0.465$\pm 0.022$ & 17 
        & 0.939  \\
        &  & x & x &  & = & 0.491$\pm 0.023$ & 16 & 0.990  \\
         $Z+X\ell n^{2}s$ & x &  &  & x & 0.583$\pm 0.077$ & 0.476$\pm 0.023$ & 17 & 0.822  \\
         & x &  & x &  & = & 0.478$\pm 0.024$ & 16 & 0.822  \\
         & \textbf{x} &  & \textbf{x} &  & = & \textbf{0.48} & \textbf{15} 
         & \textbf{0.819 } \\
        \hline
\end{tabular}}\\[2pt]
\end{table*}

\begin{table*}[t]
\caption{Results of the fits of $\sigma_{T}$ and $\rho$ data. 
The symbol = has the same meaning as in Table 1.}
\label{table:2}
\renewcommand{\arraystretch}{1.2} 
{\small
\begin{tabular}{ccccccccc}
\hline
        & \multicolumn{2}{c}{G-}   
        & \multicolumn{2}{c}{exchange}   & & 
        &  &  \\
 Model  &  \multicolumn{2}{c}{Universality}  &  \multicolumn{2}{c}{degeneracy}  
         & $\alpha_{+}$ & $\alpha_{-}$ &$N_{par}$  & $\chi^{2}/dof$  \\ 
        \cline{2-5}
         & Yes & No & Yes & No &  &  &  &   \\
        \hline
        $Xs^{\epsilon}$ &  & x &  & x & 0.66$\pm$ 0.02& 0.45$\pm$ 0.02 & 16 
        & 1.020  \\
         &  & x & x &  & = & 0.48 & 15 & 1.320  \\
        \hline  
        &  & x &  & x & 0.641$\pm 0.012$ & 0.440$\pm 0.015$ 
        & 17 & 1.024  \\
        $Z+Xs^\epsilon$ &  & x & x &  & = & 0.494$\pm 0.013$ & 16 & 1.203  \\
         & x &  &  & x & 0.602$\pm 0.014$ & 0.458$\pm 0.016$ & 17 & 0.986  \\
         & x &  & x &  & = & 0.500$\pm 0.013$ & 16 & 1.092  \\
        \hline
        &  & x &  & x & 0.816$\pm 0.001$ & 0.450$\pm 0.012$ & 16 
        & 0.941  \\
        $Z+X\ell n\ s$ &  & x & x &  & = & 0.569$\pm 0.001$ & 15 & 1.769  \\
         & x &  &  & x & 0.691$\pm 0.005$ & 0.465$\pm 0.015$ & 16 & 1.250  \\
         & x &  & x &  & = & 0.592$\pm 0.008$ & 15 & 1.944  \\
        \hline
        &  & x &  & x & 0.651$\pm 0.017$ & 0.442$\pm 0.016$ & 17 
        & 1.015  \\
        $Z+X\ell n^{2}s$ &  & x & x &  & = & 0.475$\pm 0.014$ & 16 & 1.142  \\
         & \textbf{x} &  &  & \textbf{x} & \textbf{0.552}$\bf{\pm}\textbf{ 0.048}$ 
         & \textbf{0.453}$\bf{\pm}\textbf{ 0.017}$ & \textbf{17} & \textbf{0.927}  \\
         & x &  & x &  & = & 0.457$\pm 0.015$ & 16 & 0.933  \\
        \hline
\end{tabular}}\\[2pt]
\end{table*} 

It can be seen from Tables I-II that the G-universality leads to a 
clear improvement of the description of all the considered data. 
Moreover, the G-universality leads to a mild violation of the weak 
exchange-degeneracy ($\alpha_{+}-\alpha_{-}\simeq 0.1$), in constant 
with the non-universality cases. These two independent features could 
hardly be considered as numerical accidents. It is therefore 
important to explore the validity of the 2-component G-universal 
Pomeron in all the other (non-forward) existing data.

A remarkable result is the fact that the forward data clearly favour 
the maximal Heisenberg-Froissart-Martin $\ell n^{2}s$ behaviour of 
the hadron scattering amplitude \cite{Hei65} : the absolute minimum 
of $\chi^{2}/dof$ is precisely obtained for the G-universal $\ell n^{2} s$
form of the amplitude. Our $\chi^{2}/dof$ is better than that given in 
the last edition of "Review of Particle Physics" \cite{PDG}.

Let us also note that the dipole model, corresponding to a $\ell n\ 
s$ behaviour of the scattering amplitude, has a serious pathology : 
the first component of the Pomeron $Z_{AB}$ has a \textit{negative} 
contribution to the total cross-sections. Therefore this $\ell n\ s$
fit has to be dismissed. The above pathological feature of the $\ell n\ 
s$ model was already remarked in J.R. Cudell et al. \cite{PDG}, but it 
was omitted from the "Review of Particle Physics" \cite{PDG}.

The theoretical and numerical details will be presented elsewhere
\cite{GN}.

\section*{Aknowledgements}

I thank Prof. Roberto Fiore for the kind invitation at this 
wonderfully organized meeting and Prof. Vladimir Ezhela for important 
exchanges of information during the last year. I thank Dr. Pierre 
Gauron for a careful reading of the manuscript.

\end{document}